\documentclass[3p,times,twocolumn]{elsarticle}

\usepackage{array, graphicx, subfigure}
\usepackage{ecrc}

\usepackage{graphics}

\volume{00}

\firstpage{1}

\journalname{Nuclear Physics B Proceedings Supplement}

\runauth{}

\jid{nuphbp}

\jnltitlelogo{Nuclear Physics B Proceedings Supplement}

\usepackage{amssymb}

\usepackage[figuresright]{rotating}

\begin{document}

\begin{frontmatter}


\dochead{}

\title{ $\sin^2\theta^{\rm lept}_{\rm eff}$  and  $M_W$(indirect)  extracted from  9 fb$^{-1}$  $\mu^+\mu^-$ event sample at CDF}

\author{ A. Bodek, on behalf of the CDF Collaboration}

\address{
Department of Physics and Astronomy, University of Rochester, Rochester, NY. 14627, USA\\
To be published in the proceedings of the 37th International Conference on High-Energy Physics, ICHEP 2014\\
FERMILAB-CONF-14-355-E, CDF Note 11129
}
\begin{abstract}
We report on the extraction of $\sin^2\theta^{\rm lept}_{\rm eff}$ and indirect measurement of
the mass of the W boson from the forward-backward asymmetry of $\mu^+\mu^-$
events in the $Z$ boson  mass region.  The data sample collected by the CDF detector
corresponds to the full 9 fb$^{-1}$  run II sample.  
We measure   $\sin^2 \theta^{\rm lept}_{\rm eff}  =  0.2315 \pm 0.0010$, 
 $ \sin^2 \theta_W  =  0.2233 \pm 0.0009$ and   
  $M_W ({\rm indirect})  =  80.365 \pm 0.047 \;{\rm GeV}/c^2$,
where each uncertainty includes both statistical and systematic
contributions.  Comparison with the results of the D0 collaboration is presented.
\end{abstract}
\end{frontmatter}

\section{Introduction}
Now that the Higgs mass is known, the Standard Model is over constrained. Therefore, any inconsistency between precise measurements  of SM parameters would be indicative of new physics.  The parameter that needs to be measured more precisely is  $M_W$ (with errors $<$15  MeV), or equivalently 
$\sin^2\theta_W= 1 - M_W^2/M_Z^2$ (with errors  $<$0.0003).  Similarly, in order to help resolve the  long standing  $3\sigma$ difference in   
 $\sin^2\theta^{\rm lept}_{\rm eff}(M_Z)$   between SLD and LEP,  new measurements of   $\sin^2\theta^{\rm lept}_{\rm eff}(M_Z)$ should have errors similar to SLD or  LEP ($\pm$0.0003).
 $$\sin^2\theta^{\rm lept}_{\rm eff}(LEP-1:Z pole)=0.23221\pm0.00029$$
  $$\sin^2\theta^{\rm lept}_{\rm eff}(SLD:Z pole)~~~~~~=0.23098\pm0.00026$$
 
Precise extractions of $\sin^2\theta^{\rm lept}_{\rm eff}$and  $\sin^2\theta_W= 1 - M_W^2/M_Z^2$  using the
forward-backward asymmetry  ($A_{\rm fb}$) of dilepton events produced in 
p$\bar p$ and pp collisions  are now  possible for the first time because of  three new innovations:
\begin{itemize}
\item A new technique \cite{muon-scale} for calibrating the muon energy scale
as a function of  detector  $\eta$ and $\phi$ (and sign), thus greatly reducing systematic            
errors from the energy scale.  A similar method can also used
for electrons.

\item A new  event weighting technique\cite{event-weighting}.  With this technique
all  experimental uncertainties in acceptance and efficiencies cancel (by measuring  the $\cos\theta$ coefficient $A_4$  and  using the relation
 $A_{FB}=8A_4/3$). Similarly,  additional weights can be  included for antiquark dilution, which makes the analysis independent of the acceptance in dilepton rapidity.

\item The  implementation\cite{cdf-ee} of Z fitter Effective Born Approximation (EBA) electroweak radiative corrections into the theory predictions of POWHEG and RESBOS 
 which 
 allows for a measurement of both  $\sin^2\theta^{\rm lept}_{\rm eff}$(M$_Z$) and 
  $\sin^2\theta_W= 1 - M_W^2/M_Z^2$.
 \end{itemize} 

\begin{figure*}
\centerline{\includegraphics[height=79mm,width=1.\linewidth]{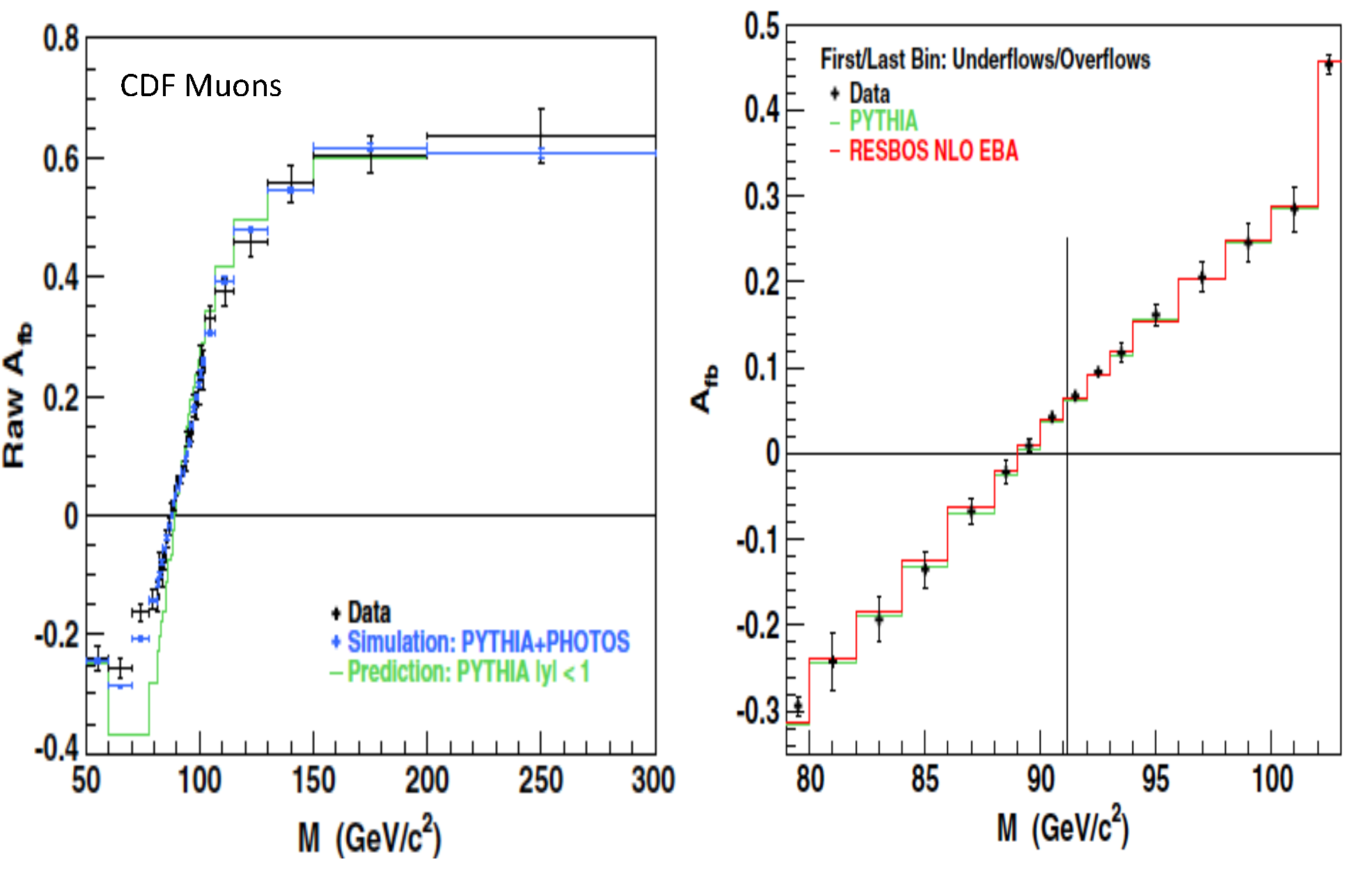}}
\caption{ Left:  CDF raw $A_{\rm fb}$ measurement in bins of $\mu^+\mu^-$
invariant mass.  Only statistical uncertainties are shown.
The Monte Carlo
simulation (\textsc{pythia}) includes the effect of resolution smearing and FSR.
The \textsc{pythia} $|y|<1$ asymmetry curve does not.
Right:  $A_{\rm fb}$ unfolded for resolution and QED-FSR.  The \textsc{pythia} calculation uses
$\sin^2\theta^{\rm lept}_{\rm eff} = 0.232$.
The EBA-based \textsc{resbos} calculation uses
$\sin^2\theta_W = 0.2233$ 
$(\sin^2\theta^{\rm lept}_{\rm eff} = 0.2315)$.$~~~~~$
}
\label{fig-raw-afb}
\end{figure*}

\subsection{Momentum-energy scale corrections}

This new technique\cite{muon-scale} is used in CDF (for both muons and electrons) and also  in CMS. In  CMS  it is used to get a precise measurement of the  Higgs mass in the four lepton channel. The technique relies on the fact that the $Z$ boson mass is well known as follows:
\begin{itemize}   
\item  Any  correlation between the scales of the two leptons is removed by getting an initial calibration using  Z events. It is done by requiring that the  mean  $\langle{1/P_T} \rangle$  of each lepton in bins of  detector $\eta$, $\phi$   and charge is equal to the expected value for generated Z events, smeared by the momentum/energy resolution.
\item    The Z mass is is used as a second order correction.  The measured  Z mass as a function of detector  $\eta$, $\phi$ and charge  of the lepton is required to be equal to the value for generated Z events (smeared by the momentum/energy resolution).
\item   Another check is the  measured  J/$\psi$ mass as a function of $\eta$ of the lepton.
\end{itemize}
The momentum/scale corrections are determined for both data events and reconstructed  hit level Monte Carlo events.
After corrections, the  reconstructed Z mass as a function $\eta$, $\phi$ and charge  for both the data and hit level  MC agree with the generator level Monte Carlo (smeared by resolution, and with experimental acceptance cuts). All charge bias is removed. 

For muons each  bin in  $\eta$ and $\phi$ the following calibration constants are
extracted.
\begin{itemize}  
\item  A multiplicative calibration correction in the quantity $1/P_T$  which accounts for possible mis-calibration of the magnetic field.
\item A calibration correction which is additive in $1/P_T$  which  accounts for tracker mis-alignments.  
\item For very low energy muons,  the J/$\psi$ mass and $\Upsilon$ mass are used to determine a small  additional calibration constant to tune the  dE/dx energy loss in the amount of material in the tracker as a function of  detector $\eta$.
\end{itemize}

 When the technique is used for electrons, the multiplicative correction accounts for tower mis-calibration and there is no additive correction since the tracker is not used in the reconstruction of the electron energy. 

\subsection{The event weighting technique}

The forward-backward  $A_{FB}$  asymmetry of leptons measured with this technique\cite{event-weighting} is insensitive to  the acceptance and lepton detection efficiency. Therefore, the raw  $A_{FB}$ which is measured using this technique is  automatically corrected for efficiency
and acceptance. The only corrections that need to be made are corrections for momentum/energy resolution which lead to event migration
between different  bins in dilepton mass.  All  experiment dependent systematic errors cancel to first order.

The event weighting technique utilizes two kinds of weights.  Angular weights are used to remove the sensitivity to acceptance and lepton detection efficiency as a function of $\cos\theta$.  In the CDF analysis, only angular weights are used.  For proton-proton collisions at the LHC, one can add weights which correct for the rapidity dependent dilution and therefore removes the sensitivity to the acceptance in Boson rapidity.


\subsection{Effective Born approximation (EBA) electroweak radiative corrections}
These radiative corrections are derived from the approach adopted at LEP\cite{cdf-ee, fitter}.
The Z-amplitude form factors are calculated by ZFITTER 6.43 \cite{fitter} which is used with LEP-1 and SLD measurement inputs for precision tests of the standard model \cite{sld-lep}.

$A_{fb}$ in the region of the mass of the $Z$ boson is  sensitive to the effective weak mixing angle
 $\sin^2\theta_{\rm eff} (M, flavor)$, where $M$ is the dilepton mass.
  Here, $sin^2\theta_{\rm eff}$ is related to the 
on-shell\cite{onshell}  electroweak mixing angle $\sin^2\theta_W= 1 - M_W^2/M_Z^2$ via complex mass and flavor dependent  electroweak radiative corrections form factors. 

 The parameter which is measured at LEP and SLD is  $\sin^2\theta^{\rm lept}_{\rm eff}$(M$_Z)$.  Previous extraction of  $\sin^2\theta^{\rm lept}_{\rm eff}$(M$_Z$) from Drell-Yan  $A_{fb}$ neglected the dependent of   $\sin^2\theta_{\rm eff}$ on flavor
and dilepton mass.  The input to the theory predictions is then one number   $\sin^2\theta_{\rm eff}$  which is assumed to be independent of mass or flavor and therefore
is interpreted as $\sin^2\theta^{\rm lept}_{\rm eff}$(M$_Z)$. 

When the full EBA  EW radiative corrections are included, the input to the theory
is $\sin^2\theta_W= 1 - M_W^2/M_Z^2$, which when compared to the data yields
a measurement of  the best fit value of $\sin^2\theta_W$. From that value of $\sin^2\theta_W$,
and the full complex EBA radiative corrections form factors one also gets the corresponding 
$\sin^2\theta^{\rm lept}_{\rm eff}(M_Z)$.  We find that 
$$\sin^2\theta^{\rm lept}_{\rm eff}(M_Z)\approx  1.037 \sin^2\theta_W.$$

\subsection{ZGRAD type EW radiative corrections}

An  approximate way to correct for the flavor dependence of  $\sin^2\theta_{\rm eff}$ from EW radiative corrections is used by the D0 collaboration. This is done by  making the following corrections  (proposed by Baur and collaborators \cite{baur}):
 $$\sin^2\theta^{\rm u-quark}_{\rm eff} =    sin^2\theta^{\rm lept}_{\rm eff}- 0.0001$$
  $$\sin^2\theta^{\rm d-quark}_{\rm eff} =    sin^2\theta^{\rm lept}_{\rm eff}- 0.0002$$

We will refer to these EW corrections as ZGRAD type corrections.  The D0 collaboration reports\cite{dzero} that  the difference between 
  $\sin^2\theta^{\rm lept}_{\rm eff}$(M$_Z)$ extracted using \textsc{resbos} (with CTEQ 6.6 -NLO PDFs) including  ZGRAD type radiative corrections and   $\sin^2\theta^{\rm lept}_{\rm eff}$(M$_Z)$ is +0.00008 larger than the value of   $\sin^2\theta^{\rm lept}_{\rm eff}$(M$_Z)$ extracted using  \textsc{pythia} 6.323 \cite{pythia3} with NNPDF2.3-NLO PDFs \cite{NNPDF}) and no EW radiative corrections. Note that   \textsc{pythia} matrix elements are QCD leading order as compared to  \textsc{resbos} matrix elements which are NLO.

The above procedure  partially corrects for the flavor dependence of $\sin^2\theta_{\rm eff}$, but does not account for the mass dependence of $\sin^2\theta_{\rm eff}$. However, since the data are dominated by events in the region of the Z boson mass, the average
 $\sin^2\theta_{\rm eff}$ is interpreted as  $sin^2\theta^{\rm lept}_{\rm eff}(M_Z)$.
Note that this  kind of analysis cannot not yield
a measurement of   $\sin^2\theta_W= 1 - M_W^2/M_Z^2$. 

\begin{figure}
\centerline{\includegraphics[height=55mm,width=1.\linewidth]{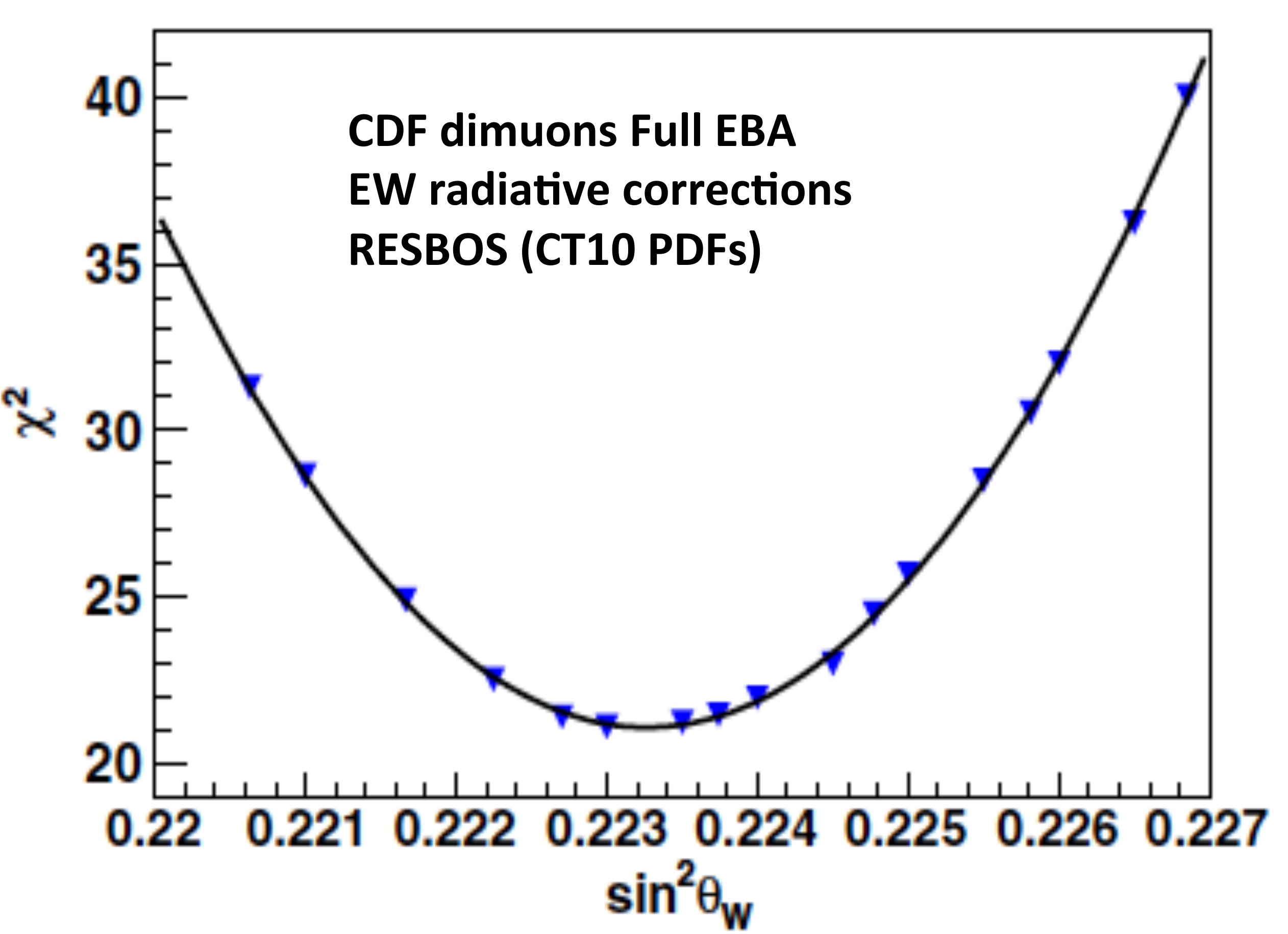}}
\caption{  $\chi^2$ comparison of the CDF $A_{\rm fb}$ $\mu^+\mu-$ measurement with \textsc{resbos}-EBA
NLO templates.   
}
\label{cdf-results}
\end{figure}

\begin{figure*}
\centerline{\includegraphics[height=85mm,width=1.\linewidth]{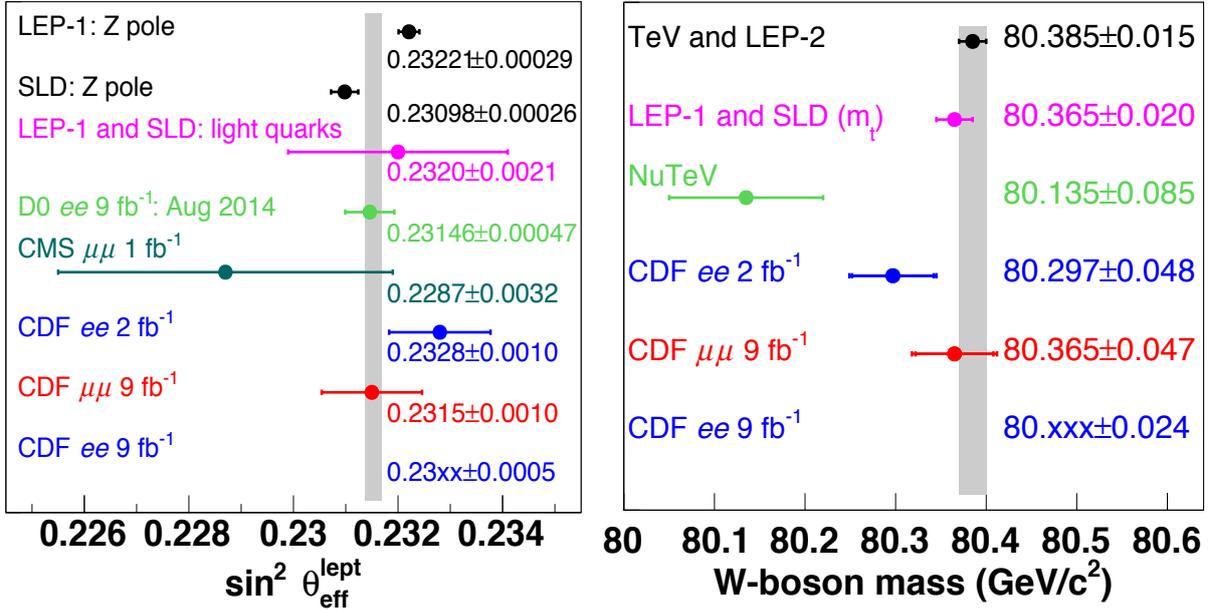}}
\caption{ Left:  Measurements of $\sin^2\theta^{\rm lept}_{\rm eff}$. Right: Direct and indirect
measurement of  $M_W$.   Also shown are the  expected errors from  9 fb$^{-1}$ $e^{+}e^{-}$ sample in CDF  (which are expected to be smaller than from the $\mu^+\mu^-$ errors by a factor of 2).
}
\label{fig-comparison}
\end{figure*}
\begin{table}
\caption{\label{tblSystErrors}
Summary of the systematic uncertainties on the extraction of
the weak mixing parameters
$\sin^2\theta^{\rm lept}_{\rm eff}$ and
$\sin^2\theta_W$.
}
\begin{tabular}{lcc}
Source  & $\sin^2\theta^{\rm lept}_{\rm eff}$ & $\sin^2\theta_W$ \\ \hline
Momentum scale & $\pm 0.00005$  & $\pm 0.00005$ \\
Backgrounds    & $\pm 0.00010$  & $\pm 0.00010$ \\
QCD scales     & $\pm 0.00003$  & $\pm 0.00003$ \\
CT10 PDFs      & $\pm 0.00037$  & $\pm 0.00036$ \\
EBA            & $\pm 0.00012$  & $\pm 0.00012$ 
\end{tabular}
\end{table}
\section{Analysis of CDF $\mu^+\mu^-$  full 9 fb$^{-1}$  run II sample}
We report on the  published analysis
of the  full 9 fb$^{-1}$  run II $\mu^+\mu^-$  data sample \cite{cdf-mumu} collected by the CDF detector.\cite{cdf-mumu}

After applying the calibrations and muon scale corrections to the experimental
and simulated data, $A_{\rm fb}$ is measured in bins
of $\mu^+\mu^-$ invariant mass using the event-weighting method.
This measurement is denoted as the raw $A_{\rm fb}$
measurement because the event-weighting method provides a first-order
acceptance correction, but does not include resolution unfolding and
final-state (FSR) QED radiation.
The raw $A_{\rm fb}$ measurement in bins of the muon-pair
invariant mass  is shown on the left part of  Fig.~\ref{fig-raw-afb}.  Only statistical uncertainties are shown.
The Monte Carlo
simulation (\textsc{pythia}+\textsc{photos}) includes the effect of resolution smearing and FSR.
To illustrate the effects of resolution smearing and FSR, the \textsc{pythia} $|y|<1$ asymmetry curve does not include the effect of resolution smearing or  FSR.

With  the event weighting technique, the events near  $\cos\theta$=0 are assigned zero weight, Therefore,
the migration of events between positive and negative $\cos\theta$ is negligible. Resolution smearing
and FSR primarily transfer events between bins in invariant mass.

The raw $A_{\rm fb}$ in bins of dimuon invariant mass is unfolded\cite{cdf-mumu} for resolution smearing and FSR using a transfer matrix
which is obtained from the Monte Carlo simulation. The unfolded $A_{\rm fb}$ is shown in the right side of Fig. \ref{fig-raw-afb}.

The electroweak (EWK) mixing parameters $\sin^2\theta^{\rm lept}_{\rm eff}$ and
$\sin^2\theta_W$ are extracted from the 
 fully unfolded 
$A_{\rm fb}$ measurements using
$A_{\rm fb}$ templates calculated with different values of
$\sin^2\theta_W$. Three  QCD 
calculations are used: LO (tree), \textsc{resbos} NLO, and
\textsc{powheg-box} NLO.  The calculations were modified
to include EWK radiative correction\cite{cdf-ee} using the
Effective Born Approximation (EBA).
For the EBA electroweak form-factor
calculations, the EW parameter is $\sin^2\theta_W$.
\par
The $A_{\rm fb}$ measurement is directly sensitive to the
effective-mixing parameters $\sin^2\theta^{\rm lept}_{\rm eff}$ which
are combinations of the form-factors and $\sin^2\theta_W$.  Most of
the sensitivity to  $\sin^2\theta^{\rm lept}_{\rm eff}$ comes from the Drell-Yan $A_{\rm fb}$
near the Z pole, where  $A_{\rm fb}$ is small.  In contrast,  $A_{\rm fb}$ at higher mass values  where  $A_{\rm fb}$
is large, is mostly sensitive to the axial coupling, which is known.

While the extracted values of the effective-mixing parameter $\sin^2\theta^{\rm lept}_{\rm eff}$
are independent of the details of the EBA model, the
interpretation of the best-fit value of $\sin^2\theta_W$ and its
corresponding form factors depend on the details
of the EBA model.

Calculations of $A_{\rm fb}(M)$ with different values of
the electroweak-mixing parameter are compared with the
measurement to determine the value of the parameter that
best describes the data. The calculations include both
quantum chromodynamic and EBA electroweak radiative corrections. 

\par
The measurement and templates are compared using the $\chi^2$
statistic evaluated with the $A_{\rm fb}$ measurement
error matrix. Each template provides a scan point for the $\chi^2$ function
$(\sin^2\theta_W, \chi^2( \sin^2\theta_W))$. The scan points
are fit to a parabolic $\chi^2$ functional form.
The $\chi^2$ distribution of the scan over templates from the
\textsc{resbos} NLO calculation (with CT10 PDFs) is shown in  
Fig.~\ref{cdf-results}.
The EBA-based \textsc{resbos} calculations of $A_{\rm fb}$
are used to extract the central value of $\sin^2\theta_W$. The other
calculations are used to estimate the systematic error from
the electroweak radiative corrections and QCD NLO radiation.

\begin{table*}[ht]
\begin{center}
\caption{\label{tblSW2results}
Extracted values of $\sin^2\theta^{\rm lept}_{\rm eff} (M_Z)$ and
$\sin^2\theta_W$ for the EBA-based QCD templates.
The \textsc{pythia} entry is the value from the scan
over non-EBA  templates calculated by  \textsc{pythia 6.4} calculations  which use the LO process matrix element with parton showering. The uncertainties of the template scans are the
measurement uncertainties ($\bar{\sigma}$). 
Entries followed by (*) are preliminary.
}
\begin{tabular}{lcccccc}
Sample &Template       & EW rad. &PDF   &$\sin^2\theta^{\rm lept}_{\rm eff}(M_Z)$ & $\sin^2\theta_W$ 	    & $\bar{\chi}^2$  \\
&(Measurement) &corr.   &  &  &       &                                \\ \hline
CDF($\mu\mu$) 9 fb$-1$&\textsc{resbos}     NLO &EBA &CT10-NLO  & $0.2315 \pm 0.0009$ & $0.2233 \pm 0.0008$ & $21.1$ \\
CDF($\mu\mu$) 9 fb$-1$&\textsc{powheg-box} NLO &EBA &CT10-NLO &$0.2314 \pm 0.0009$ & $0.2231 \pm 0.0008$ & $21.4$ \\
CDF($\mu\mu$) 9 fb$-1$&Tree LO                  & EBA& CT10-NLO&$0.2316 \pm 0.0008$ & $0.2234 \pm 0.0008$ & $24.2$ \\ \hline
CDF($\mu\mu$) 9 fb$-1$&\textsc{pythia}  6.4       &none&CTREQ5L-LO & $0.2311 \pm 0.0008$ &   $-$               & $20.8$ \\
CDF($\mu\mu$) 9 fb$-1$&\textsc{pythia}  6.4 (*)        &none&CTEQ6L1-LO  & $0.2314 \pm 0.0008$  & $-$ &  $23.6$ \\
CDF($\mu\mu$) 9 fb$-1$&\textsc{pythia}  6.4 (*)        & none &CTEQ6.6-NLO & $0.2314 \pm 0.0008$ & $-$ & $24.0$                \\ \hline
CDF($ee$) 2 fb$-1$ &A4:\textsc{resbos} NLO &EBA &CT10-NLO  &
$0.2328 \pm 0.0010$  &$0.2246 \pm 0.0009$ &  $-$ \\
 \hline
D0($ee$) 9.7 fb$-1$& \textsc{resbos}     NLO       & ZGRAD &CTEQ6.6-NLO & $0.23146 \pm 0.00047$ & $-$ & $-$ \\ 
D0($ee$) 9.7 fb$-1$&\textsc{pythia}  6.323        &none& NNPDF2.3-NLO & $0.23138 \pm 0.00047$  & $-$ &  $-$ \\
 \hline
 LEP-1  &$ -$   &$-$ &$-$  &$0.23221 \pm 0.00029$&  $-$ \\
 SLD &$ -$   &$-$ &$-$  &$0.23098 \pm 0.00026$&  $-$ \\
 LEP-1 + SLD &$ -$   &$-$ &$-$  &$0.23153 \pm 0.00016$&  $-$ \\
\hline
\end{tabular}
\end{center}
\end{table*}

\section {Systematic errors  in the extraction of   $\sin^2 \theta^{\rm lept}_{\rm eff}$  from the  full 9 fb$^{-1}$  run II sample}
In all QCD calculations, the mass-factorization and
renormalization scales are set to the muon-pair
invariant mass. To evaluate the effects of different scales,
the running scales are varied independently by a factor
ranging from $0.5$ to $2$ in the calculations.
The largest observed deviation of the best-fit value of
$\sin^2\theta_W$ from the default value is considered to be
the QCD-scale uncertainty. This uncertainty is
$\Delta\sin^2\theta_W({\rm QCD \; scale)} = \pm 0.00003$.

The CT10 PDFs are derived from a global analysis of experimental
data that utilizes 26 fit parameters and the
associated error matrix. In addition to the best global-fit
PDFs, PDFs representing the uncertainty along the eigenvectors
of the error matrix are also derived. For each eigenvector $i$,
a pair of PDFs are derived using 90\% C.L. excursions from the
best-fit parameters along its positive and negative directions.
The difference between the best-fit $\sin^2\theta_W$ values
obtained from the positive (negative) direction excursion PDF
and the global best-fit PDF
is denoted as $\delta^{+(-)}_i$. The 90\% C.L. uncertainty
for $\sin^2\theta_W$ is given by the expression
$\frac{1}{2} \sqrt{ \sum_i (|\delta^+_i|+
                            |\delta^-_i|)^2 }$,
where the sum $i$ runs over the 26 eigenvectors. This value
is scaled down by a factor of 1.645 for the 68.3\% C.L. (one
standard-deviation) uncertainty yielding
$\Delta\sin^2\theta_W({\rm PDF}) = \pm 0.00036$. The PDF error 
is expected to be a factor of 2 smaller with more modern PDFs.

\par
The \textsc{resbos} $A_{\rm fb}$ templates are the default
templates for the extraction of
$\sin^2\theta^{\rm lept}_{\rm eff}$.
The scan with the \textsc{powheg-box} or the tree templates yields
slightly different values for $\sin^2\theta_W$. The difference,
denoted as the EBA uncertainty, is
$\Delta\sin^2\theta_W({\rm EBA}) = \pm 0.00012$.
Although the \textsc{resbos} and \textsc{powheg-box} predictions
are fixed-order NLO QCD calculations at large boson $P_{\rm T}$,
they are all-orders resummation calculations in the low-to-moderate
$P_{\rm T}$ region, which provides most of the total cross
section. The EBA uncertainty is a combination of differences between
the resummation calculations and the derived value of
$\sin^2\theta_W$ with and without QCD radiation.
\par
In summary, the total systematic uncertainties on $\sin^2\theta_W$
from the QCD mass-factorization and renormalization scales, and from
the CT10 PDFs is $\pm 0.00036$.
All component uncertainties (shown in Table \ref{tblSystErrors}) are combined in quadrature. 
With the inclusion of the EBA uncertainty,
the total systematic uncertainty is $\pm 0.00038$.

\section{Summary of results from the  CDF  9 fb$^{-1}$ $\mu^+\mu^-$ sample}

The   best fit  extracted 
values of  $\sin^2\theta_{\rm eff}^{\rm lept}$, $\sin^2\theta_{W}$, and $M_W$ from the
CDF measurement of  $A_{\rm fb}$ in the  9 fb$^{-1}$ $\mu^+\mu-$ sample are:
\begin{eqnarray*}
  \sin^2 \theta^{\rm lept}_{\rm eff} & = & 0.2315 \pm 0.0010 \\
  \sin^2 \theta_W & = & 0.2233 \pm 0.0009 \\   
  M_W ({\rm indirect}) & = & 80.365 \pm 0.047 \;{\rm GeV}/c^2 \, .
\end{eqnarray*}

Each uncertainty includes  statistical errors, and various sources of systematic errors
combined in quadrature.

The results  for   $\sin^2 \theta^{\rm lept}_{\rm eff} (M_Z)$ are consistent with other
measurements at the $Z$-boson pole, as shown on the  left panel of 
Fig, \ref{fig-comparison}.  The results for $M_W$ are consistent with other direct and
indirect measurements of $M_W$ as shown on the right panel of Fig. \ref{fig-comparison}.
 
Because of the larger angular acceptance for electrons, the  error
in  $\sin^2 \theta^{\rm lept}_{\rm eff}$ for the 9 fb$^-1$ $e^{+}e^{-}$ sample
are expected to be smaller by a factor of two
 (about $\pm$0.0005). Both the statistical errors and systematic
 errors such as PDFs are smaller for events with large $\cos\theta$.
The  corresponding expected  error 
in the CDF  extracted value of    $M_W^{indirect} $ ($\pm$ 24 MeV) 
 will be competitive with the direct measurements
 of $M_W$. The results from the  CDF
   full 9 fb$^{-1}$  run II
    $e^{+}e^{-}$  sample  are expected by end of 2014.
    
    \section{Comparison of CDF and D0 results}
 Also shown in   Figure  \ref{fig-comparison} and Table \ref{tblSW2results} is the most recent (Aug. 2014)   value\cite{dzero}  of  $\sin^2 \theta^{\rm lept}_{\rm eff} (M_Z)$ extracted from the  full 9.7 fb$^{-1}$  run II $e^{+}e^{-}$  sample in D0  \cite{dzero} (0.23146 $\pm$ 0.00047).
 
 In order to make a more direct comparison with the D0 results we have done
 preliminary extractions of  $\sin^2 \theta^{\rm lept}_{\rm eff}$ from the CDF data using  
 \textsc{pythia} 6.4  \cite{pythia} with no EW radiative corrections. The values extracted with  CTEQ6.6 NLO PDFs  are the same as the values extracted with CTEQ6L1-LO PDFs as shown in Table \ref{tblSW2results}.  In contrast the values extracted using the older CTEQ5L-LO PDFs
 are 0.0003 lower.
 
We find that the values of 
  $\sin^2\theta^{\rm lept}_{\rm eff}$(M$_Z)$ extracted from CDF $\mu^+\mu^-$ data  using \textsc{resbos}(NLO) (with CT10 PDFs) including EBA radiative corrections 
  are 0.0001 larger than    $\sin^2\theta^{\rm lept}_{\rm eff}$(M$_Z)$ extracted using \textsc{pythia}  6.4 (also with CTEQ6.6-NLO PDFs) and no EW radiative corrections.
  
This difference is similar to the difference (+0.00008) 
 between   $\sin^2\theta^{\rm lept}_{\rm eff}$(M$_Z)$  extracted from the D0 $e{+}e{-}$ data using \textsc{resbos}(NLO) (with CTEQ 6.6-NLO PDFs) including  ZGRAD  type radiative corrections 
 and   $\sin^2\theta^{\rm lept}_{\rm eff}$(M$_Z)$ extracted using \textsc{pythia}  6.323 (with NNPDF2.3-NLO PDFs). Therefore, the 
CDF results (extracted with \textsc{resbos}, CT10 and EBA radiative corrections)  and the D0 
results  (extracted with \textsc{resbos}, CTEQ6.6 and ZGRAD type radiative corrections) are directly comparable.
  
When the CDF extraction of   $\sin^2\theta^{\rm lept}_{\rm eff}$(M$_Z)$ from the  full 9 fb$^{-1}$  run II $e^{+}e^{-}$ data sample  is completed, the  uncertainty in the average of  both CDF and D0   9 fb$^{-1}$  measurements
 of $\sin^2 \theta^{\rm lept}_{\rm eff}$ in the $e^{+}e^{-}$ channel  will be competitive with LEP and SLC. 


\begin{thebibliography}{00}

%
\bibitem{muon-scale} A. Bodek et al., Eur. Phys. J. C72 (2012)10 (arXiv:1208.3710). ({\it Extracting Muon Momentum Scale Corrections for Hadron Collider Experiments}).
%
\bibitem{cdf-ee} Aaltonen et. al.,  (CDF collaboration),  Phys, Rev. D88 (2013)  072002 (2013) (arXiv:1307.0770). 
({\it Indirect measurement of  $\sin^2\theta_{W}$  (or $M_W$) 
using $e^{+}e^{-}$ pairs in the Z-boson region with
 p$\bar p$ collisions at a center-of-momentum energy of 1.96 TeV})

\bibitem{event-weighting} A. Bodek, Eur. Phys. J. C67, (2010) 321 (arXiv:0911.2850).
({\it A simple event weighting technique for optimizing the measurement of the forward-backward asymmetry of Drell-Yan dilepton pairs at hadron colliders}).

\bibitem{cdf-mumu} Aaltonen et. al.,  (CDF collaboration) Phys.Rev.  D89 (2014) 072005 (arXiv:1402.2239). 
({\it Indirect measurement of  $\sin^2\theta_{W}$  (or $M_W$) using $\mu^+\mu^-$ pairs from $\gamma^*$/Z bosons produced in p$\bar p$ collisions at a center-of-momentum energy of 1.96 TeV}).

\bibitem{fitter} D. Bardin, M. Bilenky, T. Riemann, M. Sachwitz, and H. Vogt, Comput. Phys. Commun. 59, 303 (1990); D. Bardin, P. Christova, M. Jack, L. Kalinovskaya, A. Olchevski, S. Riemann, and T. Riemann, Comput. Phys. Commun. 133, 229 (2001);  A. Arbuzov, M. Awramik, M. Czakon, A. Freitas, M. Grünewald, K. Monig, S. Riemann, and T. Riemann, Comput. Phys. Commun. 174, 728 (2006).

\bibitem{sld-lep} S. Schael et al. (ALEPH, DELPHI, L3, OPAL, and SLD collaborations; LEP Electroweak Working Group; and SLD Electroweak and Heavy Flavour Groups), Phys. Rep. 427, 257 (2006) ({\it Precision electroweak measurements on the Z resonance})

\bibitem{onshell} A. Sirlin, Phys. Rev. D 22, 971 (1980).

\bibitem{baur} U. Baur, O. Brein, W. Hollik, C. Schappacher, and D. Wackeroth, Phys. Rev. D 65, 033007, 2002, arXiv:hep-ph/0108274 ({\it Electroweak Radiative Corrections to Neutral-Current Drell-Yan Processes at Hadron Colliders})

\bibitem{pythia3} T. Sjostrand, P. Ede?n, C. Feriberg, L. L?onnblad, G. Miu, S. Mrenna, and E. Norrbin, Comp. Phys. Commun. 135, 238 (2001) textsc{pythia} version v6.323 is used by D0.

\bibitem {NNPDF} R. D. Ball et al., Nucl. Phys. B867, 244 (2013) ({\it Parton distributions with LHC data, NNPDFs})

\bibitem{cteq66} P. M. Nadolsky, H. -L. Lai, Q. -H. Cao, J. Huston, J.Pumplin, D. Stump, W. -K Tung and C. -P. Yuan, Phys. Rev. D 78, 013004 (2008) ({\it  Implications of CTEQ6.6 global analysis for collider observables})

\bibitem {CT10}  H.-L. Lai, M. Guzzi, J. Huston, Z. Li, P. Nadolsky, J. Pumplin, and C.-P. Yuan, Phys. Rev. D 82, 074024 (2010) ( {\it New parton distributions CT10 for collider physics})

\bibitem{pythia} T. Sjostrand, P. Edén, L. Lönnblad, G. Miu, S. Mrenna, and E. Norrbin, Comput. Phys. Commun. 135, 238 (2001) ({\it PYTHIA 6.4 physics and manual} textsc{pythia}  6.4 is used by CDF)

\bibitem{dzero} V. M. Abazov et al. (D0 collaboration)   arXiv:1408.5016  (Aug. 2014) ({\it Measurement of the effective weak mixing angle in  p$\bar p \to  Z/\gamma^* $ $e^+e^-$ events})


\end{thebibliography}
\end{document}